\pdfoutput=1
\documentclass[3p,times,twocolumn]{elsarticle}

\usepackage[T1]{fontenc}
\usepackage{amsmath,amssymb,amsfonts}
\usepackage{booktabs}
\usepackage{graphicx}
\usepackage[hidelinks]{hyperref}
\usepackage{microtype}
\usepackage[section]{placeins}
\usepackage{siunitx}
\usepackage{algorithm}
\usepackage{algpseudocode}

\newcommand{\comm}[2]{[#1,\,#2]}

\newcommand{\dv}[2]{\frac{d#1}{d#2}}

\journal{Journal of Computational Physics}
\biboptions{numbers,sort&compress}

\begin{document}

\begin{frontmatter}

\title{Hardware-Aware Performance Characterization of Small Dense Lindblad Propagation for Near-Term Quantum Control}

\author[erau]{Rylan Malarchick\corref{cor1}}
\ead{malarchr@my.erau.edu}
\cortext[cor1]{Corresponding author}

\affiliation[erau]{organization={Department of Engineering Physics, Embry-Riddle Aeronautical University},
             city={Daytona Beach},
             state={FL},
             postcode={32114},
             country={USA}}

\begin{abstract}
Dense Lindblad propagation at the Hilbert-space sizes of near-term transmon
control ($d = 3$, $9$, $27$) sits in a regime where cache boundaries, launch
overhead, and fixed latency decide which hardware is useful, not peak
arithmetic rate. This paper characterizes that regime on two x86 CPUs, two
NVIDIA GPUs, and a low-cost FPGA. Each implementation is stated as an
algorithm in $d$ with an idealized cost built from separately measured
ceilings, and the measured gaps are attributed to specific mechanisms:
thread fork/join, the strided read of the propagator, PCI Express (PCIe)
transfers, and propagator re-upload. An end-to-end gradient-ascent (GRAPE-style) benchmark
against released QuTiP 5.2.3 shows that fast propagation does not imply
fast control optimization: the C path wins by up to two orders of magnitude
at $d = 3$ but loses at $d = 27$, where the dense LU solve inside the
Pad\'e propagator build takes over $90\%$ of the cost. The FPGA
contributes a deterministic-latency operating point, $95$ cycles per step
at a stress-characterized $108$~MHz with zero cycle spread over $1000$
trials, and its Q1.15 fixed-point drift is measured on the device. The
results are a map of operating regimes for the measured grid, with the propagator build, not propagation, as the dominant cost at the
largest size.
\end{abstract}

\begin{keyword}
Lindblad master equation \sep quantum optimal control \sep Roofline model
\sep GPU computing \sep FPGA acceleration \sep QuTiP
\end{keyword}

\end{frontmatter}

\section{Introduction}

The Lindblad master equation~\cite{Lindblad1976,Gorini1976} is the standard
Markovian model for open-quantum-system dynamics,
\begin{equation}
  \dv{\rho}{t} = -i\comm{H}{\rho}
  + \sum_k \left(L_k \rho L_k^\dagger
  - \tfrac{1}{2}\{L_k^\dagger L_k,\rho\}\right),
\end{equation}
and it appears in the main software loops of gradient-ascent pulse
engineering (GRAPE), robustness studies, and feedback-controller design for
superconducting qubits~\cite{Khaneja2005}. In
the dense, time-independent case, vectorization turns one propagation step
into
\begin{equation}
  \vec{\rho}(t+\Delta t) = P\,\vec{\rho}(t), \qquad
  P = e^{\mathcal{L}\Delta t}, \qquad
  P \in \mathbb{C}^{d^2 \times d^2}.
\end{equation}
At $d=3$, $9$, and $27$, the hot path is a 9$\times$9, 81$\times$81, or
729$\times$729 dense complex matrix--vector product (matvec). These sizes are the
ones near-term transmon control simulates: $d=3$ is one transmon truncated
to three levels, $d=9$ two coupled transmons, and $d=27$ three, the setting
of single- and few-qubit gate design with leakage.

This size regime is awkward. It is too small to inherit the intuitions of
large-matrix BLAS (basic linear algebra subprograms) libraries, but too
large for overheads to disappear. The asymptotic
arithmetic intensity of the propagation kernel approaches 0.5 FLOP/byte, so
a Roofline reading~\cite{Williams2009} says the kernel should live in the
bandwidth-sensitive part of the machine balance. In practice, however, the
relevant working sets are only about 1.5 KB, 105 KB, and 8.3 MB. That means
cache boundaries, kernel launch cost, and fixed-latency behavior matter as
much as nominal peak FLOP rate.

Existing Lindblad performance work has mostly gone in one of three
directions: general software frameworks such as QuTiP and its later
extensions~\cite{Johansson2012,Johansson2013,Lambert2026}, modern
accelerated or differentiable backends such as \texttt{qutip-jax},
\texttt{dynamiqs}, and adjacent JAX-based differential-equation software such
as Diffrax~\cite{QuTiPJax,Guilmin2024,Diffrax2026},
larger-scale parallel or algorithmic formulations~\cite{Meyerov2019}, or
hardware/control papers that assume the classical simulator as a given
rather than measuring it directly~\cite{Gebauer2020,Yang2021,Caune2024}.
Beyond that Lindblad-specific software, three adjacent bodies of work
border this study without covering it. Batched dense linear algebra addresses many small independent matrix
operations on GPUs~\cite{Abdelfattah2016,Abdelfattah2017,Dongarra2017Batched,cuBLAS},
but its batched GEMM (general matrix multiply) kernels target matrices from a few dozen to a few
hundred rows in each factor, and the control workload here is a
matrix--vector chain with one shared matrix, batched over states rather
than over matrices. Propagator- and exponential-based simulation paths,
from EXPOKIT~\cite{Sidje1998} to the action-of-the-exponential methods of
Al-Mohy and Higham~\cite{AlMohy2011}, provide the algorithmic alternatives
to the dense fixed-propagator route measured here, and
Section~\ref{sec:grape} quantifies the term such methods would remove.
Quantum optimal control software, including QuTiP's optimization stack
\cite{Lambert2026}, the Krotov package~\cite{Goerz2019Krotov}, and
integrated tool sets such as $C^3$~\cite{Wittler2021}, reports optimizer
convergence and physics results, but not hardware-resolved cost
characterizations of the inner simulation loop. What has been missing is a
controlled comparison that asks the practical question a software author
needs answered: for small dense Lindblad propagation, which hardware wins,
under which timing model, and for what notion of ``performance''?

That last point matters because isolated kernel speed is not the same thing
as end-to-end control speed. A GPU can win in device-resident timing and
still lose on wall clock because the problem is too small to hide launch and
transfer overhead. An FPGA can lose on throughput and still be the only
useful answer if the real constraint is deterministic embedded latency. A C
kernel can dominate QuTiP in propagation and still lose a GRAPE-style
application benchmark if propagator construction becomes the bottleneck.

This paper is therefore built as a comparison study. We combine a
Roofline-guided working-set analysis with measurements on two x86 CPUs, two
NVIDIA GPUs (graphics processing units), and a low-cost FPGA
(field-programmable gate array); we state each implementation as an
algorithm in $d$ and compare its measured cost with an idealized cost built
from separately measured hardware ceilings; we separate three GPU timing
models; we report deterministic FPGA cycle counts rather than USB wall
time; and we close with a GRAPE-style benchmark against released QuTiP
5.2.3. The end-to-end result reverses the kernel result at the largest
size: the C path outruns QuTiP by two orders of magnitude at $d=3$ but
loses at $d=27$, because the dense LU solve inside the Pad\'e propagator
build comes to dominate the cost of a control-optimization step. The result
is not a single universal winner. It is a map of operating regimes for the
measured grid of sizes, hosts, and timing models.

\section{Problem Formulation and Measurement Protocol}

\subsection{Kernel and Working-Set Model}

Vectorizing $\rho$ gives
\begin{equation}
  \dv{\vec{\rho}}{t} = \mathcal{L}\,\vec{\rho},
\end{equation}
with $\mathcal{L}\in\mathbb{C}^{d^2\times d^2}$. Once the propagator
$P=e^{\mathcal{L}\Delta t}$ is built, the hot path is a dense complex
matvec. Counting a complex multiply-add as eight real floating-point
operations gives
\begin{equation}
  \mathrm{FLOPs} = 8d^4, \qquad
  \mathrm{bytes} = 16(d^4 + 2d^2),
\end{equation}
so the arithmetic intensity tends to 0.5 FLOP/byte from below. The hot
working set is the propagator plus one input and one output state vector in
double precision.

\begin{table}[t]
  \centering
  \caption{Problem sizes studied in this paper.}
  \label{tab:problem_sizes}
  \small
  \begin{tabular}{@{}rrrr@{}}
    \toprule
    $d$ & $d^2$ & hot working set & arithmetic intensity \\
    \midrule
    3 & 9 & 1.5 KB & 0.41 FLOP/byte \\
    9 & 81 & 105 KB & 0.49 FLOP/byte \\
    27 & 729 & 8.3 MB & 0.50 FLOP/byte \\
    \bottomrule
  \end{tabular}
\end{table}

\subsection{Implementations Compared}

The CPU baseline is a C11 library with dense Lindbladian construction, a
Pad\'e $[13/13]$ matrix exponential~\cite{Higham2005}, and several
propagation paths: scalar array-of-structures, compiler-auto-vectorized C,
explicit AVX2 (256-bit x86 vector instructions), and OpenMP (shared-memory
threading) over independent batched states. The local
i9-13980HX configuration used in this work exposed AVX2 but not AVX-512. In
other words, the measured Linux-visible instruction set on the hybrid 8P+16E system was
AVX2-only rather than AVX-512-capable, which is consistent with the mixed
P-core/E-core configuration of this processor family. The measured comparison
is therefore scalar versus auto-vectorized versus AVX2 rather than an
AVX-512 study.

The CUDA (NVIDIA's GPU programming platform) comparison uses one thread
per output row of the dense matrix--vector product and
reports two timing models: host-visible timing, which includes packing,
transfer, launch, and copy-back, and kernel-only timing with resident device
buffers. That separation is required here because the problem is small
enough that platform overhead can dominate device arithmetic.

The FPGA comparison targets only the $d=3$ kernel. The Tang Nano 20K design
implements a 9$\times$9 complex matrix--vector product in Q1.15 fixed
point (16-bit two's complement, one sign and fifteen fractional bits), with
a UART serial link for loading and an on-chip cycle counter. The important
metric is on-chip cycle-exact latency, not USB-attached wall-clock time.

\subsection{Hardware and Fairness Choices}

The two x86 hosts were an Intel Core i9-13980HX laptop and an AMD Ryzen 5
1600 desktop. The i9 machine also hosted the RTX 4070 Laptop GPU. The Ryzen
desktop hosted the GTX 1070 Ti. The Tang Nano 20K board was attached to the
local machine for synthesis, flashing, and hardware timing.

All QuTiP timings use the pip-installable official release available during
benchmarking, \texttt{qutip==5.2.3}, installed under \texttt{uv}. The
current QuTiP documentation describes a matrix-form Lindblad path, but that
option was not available in the released 5.2.3 package installed on the
measured machines. The fair released baseline in this paper is therefore
default \texttt{mesolve}; if a later public release exposes that matrix-form
path, it should be treated as a separate baseline rather than folded into
the present results.

The main OpenMP sweep used \texttt{OMP\_PROC\_BIND=true} and
\texttt{OMP\_PLACES=cores}, but it did not use explicit \texttt{taskset} or
\texttt{numactl} pinning to force either the performance-core or
efficiency-core subset on the i9. An auxiliary scheduler-sensitivity check
with \texttt{taskset -c 0-15} (P-core set) and \texttt{taskset -c 16-31}
(E-core set) at batch size 128 found that E-core-only placement was
substantially weaker for the larger kernels, while the default scheduler
plus OpenMP core binding stayed within 4\% of explicit P-core-only
placement at $d=27$ and 16 threads. The reported i9 scaling curves are
therefore not pure P-core measurements, but their qualitative conclusions do
not depend on an unreported scheduling artifact.

\subsection{Timing Protocol}
\label{sec:protocol}

Every reported number is the median over repeated trials, and the per-trial
raw values ship as CSV files in the repository. The C application benchmark runs
five timed trials after one untimed warm-up trial. The CPU batch sweep runs
five trials of $50$ repetitions each after ten untimed batches. Both read
\texttt{CLOCK\_MONOTONIC}. The GPU benchmarks run five trials of 50
repetitions each after five warm-up repetitions; kernel-only times use CUDA
events, and host-visible and resident-$P$ times use the host monotonic
clock around the full call. The QuTiP benchmarks run five trials (twenty at
$d=3$) after one warm-up trajectory. The FPGA numbers are on-chip cycle
counts, so they need no statistics; the counter returned the same value in
every one of $1000$ trials. C and CUDA code was built with GCC 13.3.0 at
\texttt{-O3 -march=native -ffast-math} and CUDA 12.9 (i9 host) or 12.0
(Ryzen host), driver 580.173.02, Linux 7.0. QuTiP is the released
\texttt{qutip==5.2.3} wheel with NumPy 2.5.2 and SciPy 1.18.1. C and
QuTiP application benchmarks ran one after the other on an otherwise idle
host. One systematic effect deserves its own table: on the hybrid i9, QuTiP
timings vary by up to a factor of two from trial to trial under the default
BLAS threading, and pinning the process to performance cores with
single-threaded BLAS removes the spread
(Table~\ref{tab:qutip_thread_sensitivity}). Default threading is reported
as the primary number because it is what an unmodified installation does.

\subsection{Reusable Propagator Construction for the GRAPE Benchmark}

The application benchmark models one GRAPE-like landscape point as a
piecewise-constant control sequence with segment-dependent drive amplitude
$\Omega_s$. Within each segment the generator is constant, so
Eqs.~(1)--(4) apply segment by segment; the next equation introduces no new
dynamical model, only the segment generator of the benchmark. The C and
QuTiP sides simulate the same physical task: the same static Hamiltonian
$H_0 = \mathrm{diag}(0,\dots,d-1)$, the same truncated $\sigma_x$ drive,
the same two collapse operators (amplitude damping with $T_1 = 50$ and pure
dephasing with $T_2 = 30$ in the same units as $\Delta t = 0.5$), the same
initial state, and the same deterministic drive schedule, which is pinned
to identical values by unit tests on both sides. The two sides differ in
algorithm, and deliberately so: the C path builds one propagator per
segment and applies it twenty times, while released \texttt{mesolve}
integrates each segment with an adaptive solver. A second QuTiP baseline
therefore builds and applies propagators through QuTiP's own
\texttt{expm}, which matches the C algorithm and separates the effect of
the algorithm from the effect of the implementation.

The generator splits into three reusable pieces: the coherent static term,
the coherent drive term, and the dissipative term, so that each segment
forms
\begin{equation}
  \mathcal{L}(\Omega_s) = \mathcal{L}_0 + \Omega_s \mathcal{L}_{\mathrm{drv}}
  + \mathcal{L}_{\mathrm{diss}}.
\end{equation}
A reusable matrix-exponential workspace keeps repeated segment
exponentials from reallocating large temporary matrices. Unit tests verify
that the decomposed superoperator construction matches the monolithic
Lindbladian build, and that the workspace-backed propagator builder matches
the baseline propagator numerically.

\section{Algorithms and Cost Models}
\label{sec:algorithms}

This section states each measured path as an algorithm in the Hilbert-space
dimension $d$, with $n = d^2$ the length of the vectorized state, and derives
an idealized cost from hardware ceilings that were measured separately from
the kernel. Section~\ref{sec:results} then compares the idealized cost with
the measured cost and names the mechanism that explains each gap. The
algorithms do not depend on the compiler, the cache sizes, or the language.
The ceilings do, and they are listed in Tables~\ref{tab:probes_cpu}
and~\ref{tab:probes_gpu}.

\subsection{Propagator Build}

\begin{algorithm*}[t]
\caption{Propagator build, $P = e^{\mathcal{L}\Delta t}$, scaling and squaring with Pad\'e $[13/13]$~\cite{Higham2005}.}
\label{alg:build}
\begin{algorithmic}[1]
\Require $\mathcal{L} \in \mathbb{C}^{n\times n}$, step $\Delta t$, threshold $\theta$
\State $A \gets \mathcal{L}\,\Delta t$;\quad $s \gets \max(0, \lceil \log_2 (\|A\|_1 / \theta) \rceil)$;\quad $A \gets A / 2^s$
\State $A_2 \gets A A$;\; $A_4 \gets A_2 A_2$;\; $A_6 \gets A_4 A_2$ \Comment{3 products}
\State $V \gets A_6 (b_{12} A_6 + b_{10} A_4 + b_8 A_2) + b_6 A_6 + b_4 A_4 + b_2 A_2 + b_0 I$ \Comment{1 product}
\State $W \gets A_6 (b_{13} A_6 + b_{11} A_4 + b_9 A_2)$ \Comment{1 product}
\State $U \gets A \left[ W + b_7 A_6 + b_5 A_4 + b_3 A_2 + b_1 I \right]$ \Comment{1 product}
\State Solve $(V - U)\, P = (V + U)$ for $P$ by LU with partial pivoting \Comment{dense $\tfrac{2}{3} n^3$}
\For{$i = 1$ \textbf{to} $s$} $P \gets P P$ \EndFor \Comment{$s$ products}
\end{algorithmic}
\end{algorithm*}

Algorithm~\ref{alg:build} is the standard Higham construction. Two choices
in this implementation matter for cost. First, the threshold is
$\theta = 0.25$ instead of Higham's $5.37$, which raises $s$ by about four.
Higham's value bounds the backward error of the approximant. This
single-degree implementation has no degree selection, and at
$\|\mathcal{L}\Delta t\|_1 \approx 13$ (the $d = 27$ generator) the
forward error at the larger threshold reached $5\times10^{-7}$. The smaller
threshold brings it to about $10^{-9}$ for the small dense matrices and
$10^{-13}$ for the larger superoperators. Second, the product routine skips zero
entries of the left factor, so the six Pad\'e products and the $s$
squarings run in time proportional to the number of nonzeros of the
powers of $A$, which stay sparse for the diagonal-plus-drive Hamiltonians
and few collapse operators of this problem. The LU solve has no such
shortcut. For $n = 729$ it performs $\tfrac{2}{3} n^3 \approx 2.6\times10^8$
complex operations on a dense matrix, and Section~\ref{sec:grape} shows that
this one line is $92\%$ to $94\%$ of the build time at $d = 27$.

\subsection{Propagation Step}

\begin{algorithm}[t]
\caption{Propagation step, $\vec\rho_{\mathrm{out}} = P\,\vec\rho_{\mathrm{in}}$, row-major dense complex matvec.}
\label{alg:step}
\begin{algorithmic}[1]
\Require $P \in \mathbb{C}^{n\times n}$ row-major, $\vec\rho_{\mathrm{in}} \in \mathbb{C}^{n}$
\For{$i = 0$ \textbf{to} $n-1$}
  \State $a \gets 0$
  \For{$j = 0$ \textbf{to} $n-1$} $a \gets a + P_{ij}\,\rho_j$ \EndFor
  \State $\rho^{\mathrm{out}}_i \gets a$
\EndFor
\end{algorithmic}
\end{algorithm}

Algorithm~\ref{alg:step} is the hot loop of every platform. One step
performs $n^2$ complex multiply-adds, which is $8n^2 = 8d^4$ real
floating-point operations, and its compulsory traffic is one read of $P$,
one read of $\vec\rho_{\mathrm{in}}$, and one write of
$\vec\rho_{\mathrm{out}}$, $16(n^2 + 2n)$ bytes in double precision. The
density matrix is stored row-major, so the $d\times d$ array is already the
$n$-vector and no reshaping copy exists. On the CPU the compiler emits one
fused multiply-add per two complex elements into a single accumulator, so the
inner loop is a dependency chain of length $n/2$ with latency $L_{\mathrm{fma}}$
cycles per link.

\subsection{CPU: Batched MIMD over Independent States}

\begin{algorithm*}[t]
\caption{Batched CPU propagation, $T$ threads, batch of $b$ independent states.}
\label{alg:cpu}
\begin{algorithmic}[1]
\Require shared $P$, private $\vec\rho^{(k)}_{\mathrm{in}}, \vec\rho^{(k)}_{\mathrm{out}}$ for $k = 0,\dots,b-1$
\State fork $T$ threads \Comment{cost $t_{\mathrm{fork}}$ once per call}
\ForAll{$k$ in static chunks of $\lceil b/T \rceil$ per thread, in parallel}
  \State Algorithm~\ref{alg:step} on $(P, \vec\rho^{(k)}_{\mathrm{in}}, \vec\rho^{(k)}_{\mathrm{out}})$
\EndFor
\State join
\end{algorithmic}
\end{algorithm*}

Algorithm~\ref{alg:cpu} is the MIMD (multiple instruction, multiple data)
path: it parallelizes only across states. The time steps of one
state stay sequential, so this is the only parallelism the problem offers
without changing the algorithm. $P$ is read by every thread and is never
copied. The idealized cost per state-step is
\begin{equation}
  \begin{aligned}
  T_{\mathrm{cpu}}(n, T, b) ={}&
  \max\!\left(\frac{16 n^2}{B_T(16n^2)},\;
              \frac{n^2}{2}\,\frac{L_{\mathrm{fma}}}{f}\right)
  \frac{\lceil b/T \rceil}{b} \\
  &+ \frac{t_{\mathrm{fork}}(T)}{b},
  \end{aligned}
  \label{eq:cost_cpu}
\end{equation}
where $B_T(W)$ is the per-thread read bandwidth when $T$ threads stream a
shared buffer of $W$ bytes, $f$ is the core clock, and
$t_{\mathrm{fork}}(T)$ is the cost of one empty parallel region. The first
term is the Roofline bandwidth bound with the ceiling taken at the cache
level that holds $P$. The second is the single-accumulator latency bound. No constant is fitted to the measured kernel times. The two terms are
idealized estimates rather than strict bounds: the ceilings are themselves
measured with a simple probe kernel, and the propagation kernel can exceed
a probe rate by tens of percent through hardware prefetching, so ratios
slightly below one occur. Loop overhead per row (a horizontal reduction, a
scalar tail for odd $n$, one store) is left out on purpose, so that its
size appears as the gap at small $n$ rather than being absorbed.

\subsection{GPU: One Thread per Output Element}

\begin{algorithm}[t]
\caption{GPU propagation, $b$ states, one thread per $(k, i)$.}
\label{alg:gpu}
\begin{algorithmic}[1]
\Require $P$ and packed states resident in device memory
\State launch $b\,n$ threads in blocks of 256 \Comment{cost $t_{\mathrm{launch}}$}
\ForAll{threads $(k, i)$ in parallel}
  \State $a \gets 0$
  \For{$j = 0$ \textbf{to} $n-1$} $a \gets a + P_{ij}\,\rho^{(k)}_j$ \EndFor \Comment{$P_{ij}$: stride-$n$ gather across the warp}
  \State $\rho^{(k),\mathrm{out}}_i \gets a$ \Comment{coalesced store}
\EndFor
\end{algorithmic}
\end{algorithm}

In Algorithm~\ref{alg:gpu} consecutive threads of a warp own consecutive rows
$i$, so at inner index $j$ they read $P_{ij}$ at addresses $n$ elements apart.
That gather is the main departure from the idealized access pattern, and the
probe in Table~\ref{tab:probes_gpu} measures its cost directly. The state
vector is a broadcast read and the store is coalesced. Three timing models
apply, from innermost to outermost:
\begin{align}
  T_{\mathrm{kernel}} &= \frac{t_{\mathrm{launch}} + 16 n^2 b / B_{\mathrm{dev}}}{b},
  \label{eq:cost_kernel}\\
  T_{\mathrm{resident}} &= T_{\mathrm{kernel}}
  + \frac{t_{\mathrm{h2d}}(16nb) + t_{\mathrm{d2h}}(16nb) + t_{\mathrm{sync}}}{b},
  \label{eq:cost_resident}\\
  T_{\mathrm{host}} &= T_{\mathrm{resident}}
  + \frac{t_{\mathrm{h2d}}(16n^2) + t_{\mathrm{alloc}}}{b},
  \label{eq:cost_host}
\end{align}
where $B_{\mathrm{dev}}$ is the best coalesced streaming rate the device
reached in the probe, $t_{x}(m) = t_{\mathrm{lat}} + m / B_{\infty}$ is a
transfer of $m$ bytes over PCIe with latency and asymptotic bandwidth from
the probe, and $t_{\mathrm{alloc}}$ covers three \texttt{cudaMalloc} and
three \texttt{cudaFree} calls at a stated $3\,\mu$s each. Kernel-only timing
is what a device-resident simulation loop would see. Resident-$P$ timing is a
control loop that keeps the propagator on the device and moves only states.
Host-visible timing, the model used in the first version of this paper,
re-uploads $P$ and allocates on every call, and Eq.~\eqref{eq:cost_host}
shows why it carries a $16 n^2 / b$ term that the other two do not.

\subsection{FPGA: Element-Serial Fixed-Point Pipeline}

\begin{algorithm}[t]
\caption{FPGA propagation step, $k$ complex multiply-accumulate units, $Q1.15$ fixed point.}
\label{alg:fpga}
\begin{algorithmic}[1]
\Require $P$ in block RAM ($2 \times 16 n^2$ bits), $\vec\rho$ and 42-bit accumulators in registers
\For{each $(i, j)$ in row-major order, $k$ per cycle}
  \State $\mathrm{acc}_i \gets \mathrm{acc}_i + P_{ij}\,\rho_j$ \Comment{4 real multiplies}
\EndFor
\State drain the pipeline \Comment{4 cycles}
\For{$i = 0$ \textbf{to} $n-1$} $\rho_i \gets \mathrm{sat}(\lfloor \mathrm{acc}_i / 2^{15} \rfloor)$ \EndFor \Comment{$n$ cycles}
\end{algorithmic}
\end{algorithm}

Algorithm~\ref{alg:fpga} has no cache and no scheduler, so its cost is a
cycle count,
\begin{equation}
  C_{\mathrm{fpga}}(n, k) = \frac{n^2}{k} + n + 4,
  \qquad
  C(N) = N\,C_{\mathrm{fpga}} + 1,
  \label{eq:cost_fpga}
\end{equation}
in cycles per step and cycles for $N$ chained steps, with $k = 1$ in the
measured design. For $n = 9$ that is $94$ cycles per step
and $94N + 1$ for a run, which is what the on-chip counter reports
(Section~\ref{sec:fpga}). The design generalizes in $d$ by widening three
counters, the accumulator to $33 + \lceil \log_2 n \rceil$ bits, and the
block-RAM depth. The block-RAM bound is the practical limit: $P$ needs
$32 n^2$ bits, which is $26$~KB at $d = 9$ and $2.1$~MB at $d = 27$. The
Tang Nano 20K holds $828$~Kbit, so this part can host $d = 9$ but not
$d = 27$.

\subsection{Ceilings Used by the Models}

\begin{table}[t]
  \centering
  \caption{CPU ceilings from the stream probe: single-thread read bandwidth in
  GB/s at each working-set size, the aggregate rate when $T$ threads read one
  shared 8.3~MB buffer, and the cost of one empty OpenMP parallel region.
  Medians of five trials.}
  \label{tab:probes_cpu}
  \scriptsize
  \resizebox{\columnwidth}{!}{\begin{tabular}{@{}lrrrrrr@{}}
\toprule
host & 1.5 KB & 105 KB & 8.3 MB & 64 MB & shared 8.3 MB, $T$ threads & fork/join ($\mu$s) \\
\midrule
i9-13980HX & 101 & 90 & 25 & 17 & 169 (8) & 13.6 (8) \\
Ryzen 5 1600 & 92 & 77 & 31 & 24 & 128 (6) & 1.5 (6) \\
\bottomrule
\end{tabular}
}
\end{table}

\begin{table*}[t]
  \centering
  \caption{GPU ceilings from the device probe: launch latency, launch plus
  synchronize, small-copy latency, asymptotic PCIe bandwidth with pageable
  host memory host-to-device (H2D) and device-to-host (D2H), and device
  read bandwidth in GB/s for a coalesced stream and for the stride-$n$
  gather of Algorithm~\ref{alg:gpu}. Medians of five trials.}
  \label{tab:probes_gpu}
  \scriptsize
  \resizebox{\textwidth}{!}{\begin{tabular}{@{}lrrrrrrrrr@{}}
\toprule
GPU & launch ($\mu$s) & launch+sync ($\mu$s) & copy latency ($\mu$s) & H2D (GB/s) & D2H (GB/s) & stream 8.3 MB & gather 8.3 MB & stream 64 MB & gather 64 MB \\
\midrule
RTX 4070 & 1.4 & 4.7 & 4.2 & 10.9 & 10.3 & 644 & 85 & 240 & 192 \\
GTX 1070 Ti & 1.9 & 5.0 & 4.0 & 11.5 & 10.1 & 211 & 46 & 221 & 80 \\
\bottomrule
\end{tabular}
}
\end{table*}

Tables~\ref{tab:probes_cpu} and~\ref{tab:probes_gpu} hold every measured
constant the models use. Two of them decide the shape of the results. The
fork/join cost on the i9 is $13.6\,\mu$s at $8$ threads, against $1.5\,\mu$s
on the Ryzen at $6$, because the hybrid part schedules threads across
performance and efficiency cores. That single number is why threading cannot
help the $d = 3$ kernel on the i9: one batch of $128$ states at $d = 3$ takes
$11\,\mu$s of arithmetic. On the GPUs, the gather pattern of
Algorithm~\ref{alg:gpu} reads $P$ at $85$~GB/s on the RTX 4070 where a
coalesced stream of the same bytes reaches $644$~GB/s, a factor of $7.6$, and
at $46$ against $211$~GB/s on the GTX 1070 Ti. Three constants are assumptions, not measurements: the FMA latencies
$L_{\mathrm{fma}} = 4$ (i9) and $5$ (Ryzen) cycles come from the vendors'
instruction tables; the sustained single-core clocks $f = 5.0$~GHz (i9)
and $3.4$~GHz (Ryzen) are assumed values below each part's rated boost
clock, since the clock during the runs was not logged; and
$t_{\mathrm{alloc}}$ is a stated $3\,\mu$s per allocation call.

\section{Results}
\label{sec:results}

\subsection{Empirical Roofline Summary}

The empirical Roofline summary is easier to read as direct numbers because
the arithmetic intensities in Table~\ref{tab:problem_sizes} barely move
across $d=3$, $9$, and $27$. Table~\ref{tab:roofline_summary} therefore
reports, for each platform, the best bandwidth and compute rates that
\emph{this kernel} attained anywhere on the tested configuration grid,
next to the best measured point at each $d$, in the Roofline spirit of
Williams \textit{et al.}~\cite{Williams2009}. These attained ceilings are
properties of the kernel on the machine, not hardware limits; the
kernel-independent ceilings measured by the probes are in
Tables~\ref{tab:probes_cpu} and~\ref{tab:probes_gpu}. The CPU rows show
the cache-sensitive small-kernel regime directly. The GPU rows show the
resident-kernel opportunity directly, but only under kernel-only timing.

\begin{table*}[t]
  \centering
  \caption{Empirical Roofline summary expressed as direct numbers. The
  ceiling columns are the best rates this kernel attained anywhere on the
  tested configuration grid, not hardware limits; kernel-independent
  ceilings are in Tables~\ref{tab:probes_cpu} and~\ref{tab:probes_gpu}.
  The $d=3$, $d=9$, and $d=27$ columns report the best measured point
  performance at fixed $d$. GPU rows use kernel-only timing.}
  \label{tab:roofline_summary}
  \small
  \resizebox{\textwidth}{!}{\begin{tabular}{@{}lrrrrr@{}}
\toprule
platform & attained BW ceiling (GB/s) & attained compute ceiling (GFLOP/s) & $d=3$ point & $d=9$ point & $d=27$ point \\
\midrule
i9-13980HX & 319.5 & 159.3 & 14.5 & 87.8 & 159.3 \\
Ryzen 5 1600 & 120.3 & 58.7 & 26.2 & 58.7 & 47.2 \\
RTX 4070 (kernel-only) & 353.1 & 176.1 & 13.4 & 103.9 & 176.1 \\
GTX 1070 Ti (kernel-only) & 335.3 & 167.2 & 14.9 & 126.4 & 167.2 \\
\bottomrule
\end{tabular}
}
\end{table*}

The important difference is not a big change in arithmetic intensity; it is
the change in attainable ceiling. The dense matvec keeps roughly the same
intensity across the three problem sizes, but the attainable performance
still depends strongly on whether the working set is living in cache or
spilling deeper into the hierarchy. A single headline number would hide
that dependence, so the following subsections report each regime
separately.

\subsection{Idealized Versus Measured}
\label{sec:cost_model}

\begin{table}[t]
  \centering
  \caption{Idealized cost from Eq.~\eqref{eq:cost_cpu} and
  Eq.~\eqref{eq:cost_fpga} against the measured median, in ns per
  state-step, with the mechanism that explains the ratio. CPU rows at
  $d = 4$ and $d = 8$ are the spot check requested for sizes between the
  primary grid.}
  \label{tab:cost_model_cpu}
  \scriptsize
  \resizebox{\columnwidth}{!}{\begin{tabular}{@{}lrrrrrrl@{}}
\toprule
platform & $d$ & threads & batch & idealized (ns) & measured (ns) & ratio & gap mechanism \\
\midrule
i9 & 3 & 1 & 1 & 32 & 311 & 9.60 & loop overhead per row \\
i9 & 9 & 1 & 1 & 2624 & 3351 & 1.28 & FMA chain \\
i9 & 27 & 1 & 1 & 344337 & 293030 & 0.85 & bandwidth \\
i9 & 3 & 8 & 128 & 110 & 83 & 0.76 & loop overhead per row \\
i9 & 9 & 8 & 128 & 434 & 598 & 1.38 & FMA chain + fork/join \\
i9 & 27 & 8 & 128 & 50334 & 53194 & 1.06 & bandwidth + fork/join \\
i9 & 3 & 24 & 128 & 239 & 298 & 1.25 & loop overhead per row \\
i9 & 9 & 24 & 128 & 477 & 682 & 1.43 & bandwidth + fork/join \\
i9 & 27 & 24 & 128 & 25532 & 29613 & 1.16 & bandwidth + fork/join \\
i9 & 4 & 1 & 128 & 102 & 116 & 1.13 & loop overhead per row \\
i9 & 8 & 1 & 128 & 1638 & 1813 & 1.11 & FMA chain \\
i9 & 4 & 8 & 128 & 119 & 96 & 0.81 & loop overhead per row \\
i9 & 8 & 8 & 128 & 311 & 453 & 1.46 & FMA chain + fork/join \\
Ryzen & 3 & 1 & 1 & 60 & 468 & 7.87 & loop overhead per row \\
Ryzen & 9 & 1 & 1 & 4824 & 5461 & 1.13 & FMA chain \\
Ryzen & 27 & 1 & 1 & 390765 & 507432 & 1.30 & FMA chain \\
Ryzen & 3 & 6 & 128 & 22 & 25 & 1.11 & loop overhead per row \\
Ryzen & 9 & 6 & 128 & 841 & 894 & 1.06 & FMA chain + fork/join \\
Ryzen & 27 & 6 & 128 & 68433 & 94051 & 1.37 & bandwidth + fork/join \\
Ryzen & 3 & 12 & 128 & 29 & 38 & 1.31 & loop overhead per row \\
Ryzen & 9 & 12 & 128 & 438 & 1053 & 2.40 & FMA chain + fork/join \\
Ryzen & 27 & 12 & 128 & 37954 & 90073 & 2.37 & bandwidth + fork/join \\
Tang Nano & 3 & 1 & 1 & 880 & 880 & 1.00 & exact: counter model \\
Tang Nano & 3 & 1 & 1000 & 870 & 870 & 1.00 & exact: counter model \\
\bottomrule
\end{tabular}
}
\end{table}

\begin{table*}[t]
  \centering
  \caption{Idealized cost from Eqs.~\eqref{eq:cost_kernel}
  to~\eqref{eq:cost_host} against the measured median for each GPU timing
  model, in ns per state-step. A ratio above one means the device fell short
  of the coalesced-stream ideal, through too few threads at small $bn$ or
  through the stride-$n$ gather of $P$. A ratio below one means the transfer
  latency in the model is pessimistic for that size.}
  \label{tab:cost_model_gpu}
  \scriptsize
  \resizebox{\textwidth}{!}{\begin{tabular}{@{}lrrrrrrrrrrr@{}}
\toprule
 & & & \multicolumn{3}{c}{kernel-only} & \multicolumn{3}{c}{resident $P$} & \multicolumn{3}{c}{host-visible} \\
\cmidrule(lr){4-6} \cmidrule(lr){7-9} \cmidrule(lr){10-12}
GPU & $d$ & batch & idealized & meas. & ratio & idealized & meas. & ratio & idealized & meas. & ratio \\
\midrule
RTX 4070 & 3 & 1 & 1374 & 3731 & 2.7 & 14047 & 11788 & 0.8 & 36374 & 83357 & 2.3 \\
RTX 4070 & 3 & 8 & 173 & 516 & 3.0 & 1781 & 1590 & 0.9 & 4572 & 10555 & 2.3 \\
RTX 4070 & 3 & 32 & 45 & 193 & 4.3 & 467 & 494 & 1.1 & 1165 & 2671 & 2.3 \\
RTX 4070 & 3 & 128 & 13 & 48 & 3.8 & 139 & 164 & 1.2 & 313 & 709 & 2.3 \\
RTX 4070 & 9 & 1 & 1535 & 14598 & 9.5 & 14426 & 22988 & 1.6 & 46260 & 110734 & 2.4 \\
RTX 4070 & 9 & 8 & 335 & 4194 & 12.5 & 2160 & 6044 & 2.8 & 6140 & 15972 & 2.6 \\
RTX 4070 & 9 & 32 & 206 & 1048 & 5.1 & 846 & 1736 & 2.1 & 1841 & 4383 & 2.4 \\
RTX 4070 & 9 & 128 & 174 & 505 & 2.9 & 518 & 996 & 1.9 & 766 & 1619 & 2.1 \\
RTX 4070 & 27 & 1 & 14577 & 280416 & 19.2 & 29428 & 296424 & 10.1 & 831336 & 1364090 & 1.6 \\
RTX 4070 & 27 & 8 & 13377 & 35153 & 2.6 & 17162 & 40442 & 2.4 & 117401 & 172898 & 1.5 \\
RTX 4070 & 27 & 32 & 13248 & 26171 & 2.0 & 15848 & 29973 & 1.9 & 40908 & 64381 & 1.6 \\
RTX 4070 & 27 & 128 & 13216 & 24144 & 1.8 & 15520 & 27272 & 1.8 & 21785 & 37469 & 1.7 \\
GTX 1070 Ti & 3 & 1 & 1946 & 4650 & 2.4 & 14971 & 15846 & 1.1 & 37061 & 190531 & 5.1 \\
GTX 1070 Ti & 3 & 8 & 248 & 593 & 2.4 & 1900 & 1997 & 1.1 & 4661 & 23926 & 5.1 \\
GTX 1070 Ti & 3 & 32 & 66 & 175 & 2.6 & 499 & 549 & 1.1 & 1190 & 5988 & 5.0 \\
GTX 1070 Ti & 3 & 128 & 21 & 44 & 2.1 & 149 & 158 & 1.1 & 322 & 1534 & 4.8 \\
GTX 1070 Ti & 9 & 1 & 2414 & 12622 & 5.2 & 15653 & 23742 & 1.5 & 46726 & 212681 & 4.6 \\
GTX 1070 Ti & 9 & 8 & 717 & 2536 & 3.5 & 2582 & 4202 & 1.6 & 6466 & 27989 & 4.3 \\
GTX 1070 Ti & 9 & 32 & 535 & 638 & 1.2 & 1182 & 1271 & 1.1 & 2153 & 7373 & 3.4 \\
GTX 1070 Ti & 9 & 128 & 489 & 415 & 0.8 & 832 & 827 & 1.0 & 1074 & 2476 & 2.3 \\
GTX 1070 Ti & 27 & 1 & 40339 & 230949 & 5.7 & 55504 & 244429 & 4.4 & 814187 & 1584606 & 1.9 \\
GTX 1070 Ti & 27 & 8 & 38641 & 41287 & 1.1 & 42433 & 46470 & 1.1 & 137268 & 220965 & 1.6 \\
GTX 1070 Ti & 27 & 32 & 38459 & 25430 & 0.7 & 41032 & 28952 & 0.7 & 64741 & 75215 & 1.2 \\
GTX 1070 Ti & 27 & 128 & 38414 & 45908 & 1.2 & 40682 & 48652 & 1.2 & 46609 & 64697 & 1.4 \\
\bottomrule
\end{tabular}
}
\end{table*}

Tables~\ref{tab:cost_model_cpu} and~\ref{tab:cost_model_gpu} put the
idealized costs of Section~\ref{sec:algorithms} beside the measurements.
On the CPU the model lands within a factor $1.5$ at every point with
$d \ge 4$ and up to one thread per physical core, and the active term
changes with $d$: the FMA chain at $d = 8$ and $d = 9$, the L3 bandwidth at
$d = 27$ on the i9, and the FMA chain again at $d = 27$ on the
slower-clocked Ryzen. The single-state $d = 3$ kernel is $8$ to $10$ times
slower than either estimate. Its $311$~ns on the i9 is $35$~ns per row of
nine elements, which is the per-row reduction, tail, and store of
Algorithm~\ref{alg:step}, not the arithmetic. That overhead is amortized
inside a batch, and the batched $d = 3$ rows are within $25\%$ of the
model. At the full thread count the i9 stays within $1.5$ ($24$ threads),
while the Ryzen at $12$ threads reaches $2.4$ at $d = 9$ and $d = 27$:
its $12$ threads are $6$ cores with simultaneous multithreading, so two
threads share one FMA unit and the per-thread latency term of
Eq.~\eqref{eq:cost_cpu} is optimistic by about that factor. On the GPUs the
kernel-only ratios show two regimes. Below about $4{,}000$ threads
($bn < 4096$) the device is underfilled and the ratio grows with $n$, up to
$19$ at $d = 27$, $b = 1$, where each of the $729$ threads walks its row
alone. Once the device is filled the ratio is $1.8$ to $2.9$ on the RTX
4070 and $0.7$ to $1.2$ on the GTX 1070 Ti, which is the gather penalty of
Table~\ref{tab:probes_gpu} after the L2 cache absorbs part of it. With
$b \ge 8$ the resident-$P$ estimate is within $0.9$ to $2.8$ of the
measurement on the RTX 4070 and $0.7$ to $1.6$ on the GTX 1070 Ti, and the
host-visible estimate within $1.5$ to $2.6$ and $1.2$ to $5.1$; the widest
host-visible gaps are at $d = 3$ on the older GPU, where the assumed
allocation cost is the whole estimate. The FPGA row is exact: the counter
model of Eq.~\eqref{eq:cost_fpga} reproduces every measured count.

\subsection{CPU: Threading Helps Only When Independent Work Exists}

The CPU result is not ``use all cores''; it is ``batch only when the work is
actually independent.'' At $d=3$, the working set is small enough that
single-state propagation is already cache-resident and OpenMP overhead hurts
more than it helps. The best measured single-state latencies were 309.5
ns/step on the i9 and 457.6 ns/step on the Ryzen.

That changes once the problem is large enough and the batch is large enough.
At batch size 128, the i9 reaches 87.8 GFLOP/s at $d=9$ and 159.3 GFLOP/s at
$d=27$. The Ryzen peaks at 58.7 and 47.2 GFLOP/s, respectively. The newer
Intel host benefits from a larger cache hierarchy and a stronger memory
system, while the older Ryzen still provides a useful lower-end x86 point.

\begin{figure*}[t]
  \centering
  \includegraphics[width=\textwidth]{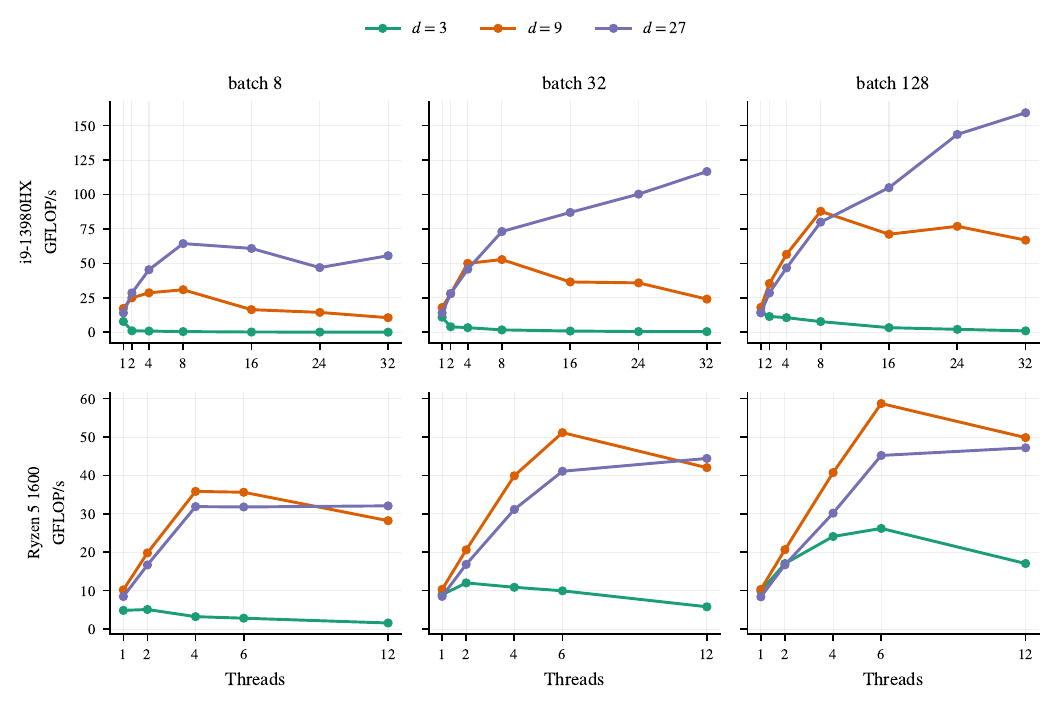}
  \caption{CPU thread scaling at batch sizes 8, 32, and 128 (columns) on
  both hosts (rows). Colors encode $d$ and the legend is shared. Threading
  does not help the tiny $d=3$ kernel at any measured batch, while $d=9$
  and $d=27$ benefit once the batch supplies enough independent work; the
  fork/join costs of Table~\ref{tab:probes_cpu} set where that happens.}
  \label{fig:cpu_threads}
\end{figure*}

\begin{table}[t]
  \centering
  \caption{Representative CPU summary. The $d=3$ entry uses the best measured
  single-state latency. The $d=9$ and $d=27$ entries use the peak measured
  throughput over the tested thread and batch grid.}
  \label{tab:cpu_summary}
  \small
  \resizebox{\columnwidth}{!}{\begin{tabular}{@{}lrrr@{}}
\toprule
host & $d=3$ latency (ns/step) & $d=9$ peak (GFLOP/s) & $d=27$ peak (GFLOP/s) \\
\midrule
i9-13980HX & 309.5 & 87.8 & 159.3 \\
Ryzen 5 1600 & 457.6 & 58.7 & 47.2 \\
\bottomrule
\end{tabular}
}
\end{table}

Figure~\ref{fig:cpu_threads} states the thread-scaling result within the
limits of the data: parallel gains in this problem attach to independent
batched states, not to the time steps of one trajectory, and appear only
once the batch amortizes the fork/join cost of
Table~\ref{tab:probes_cpu}.

\subsection{GPU: Kernel-Only Wins Appear Before Host-Visible Wins}

The GPU result depends on the timing model. In the current CUDA
implementation, host-visible timing never beats the best tested CPU across
the measured $(d,\mathrm{batch})$ grid. Kernel-only timing does cross the
CPU at larger batches for $d=9$ and $d=27$, which means the missing
wall-clock speedup is overhead, not device arithmetic. The resident-$P$
model of Eq.~\eqref{eq:cost_resident} sits between the two
(Table~\ref{tab:cost_model_gpu}): keeping the propagator on the device
removes the dominant $16n^2/b$ upload term of the host-visible path, and
what remains is the PCIe round trip for the states plus synchronization.

The two GPU generations do not order simply. The RTX 4070 does not lead
the GTX 1070 Ti at every operating point: for some of the smaller
kernel-only cases, especially at $d=9$, the older Pascal GPU matches or
beats it in the current implementation. The RTX 4070 leads only once the
larger $d=27$, larger-batch operating points amortize the fixed costs.

\begin{figure*}[t]
  \centering
  \includegraphics[width=\textwidth]{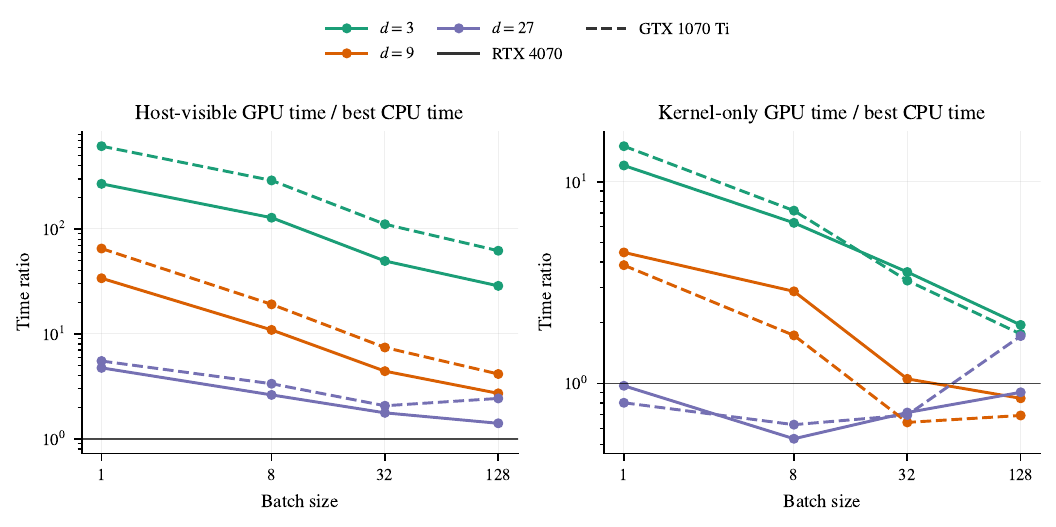}
  \caption{GPU time normalized by the best tested CPU time at the same
  $(d,\mathrm{batch})$. Colors encode $d$ and line style encodes GPU model,
  with one legend shared across both panels. A value below 1 means the GPU
  is faster. Host-visible timing never crosses below 1 in the measured
  range. Kernel-only timing does, which shows that the device math can win
  before the full platform path does.}
  \label{fig:gpu_ratio}
\end{figure*}

Figure~\ref{fig:gpu_ratio} summarizes the GPU result. The host-visible
panel never crosses 1. The kernel-only panel does. The two timing models
answer different deployment questions, and collapsing them into one number
would discard the distinction the measurement exists to make.

\subsection{FPGA: Deterministic Latency Rather Than Throughput}
\label{sec:fpga}

The FPGA result is about deterministic latency, not throughput. The Tang
Nano design targets only $d=3$, but it contributes something the CPU and
GPU sections cannot. The design runs at 108~MHz, uses 1035 four-input lookup tables (LUT4), 4
digital-signal-processing (DSP) multipliers, and 2 block RAMs, and matches the software Q1.15 reference bit
for bit over single-step and multi-step tests. The cycle counter measured a
95-cycle single step, 879.6~ns, with zero spread over 1000 trials.
Multi-step timing followed $94n+1$ cycles, the counter model of
Eq.~\eqref{eq:cost_fpga}: a one-cycle start around a 94-cycle steady-state
pipeline.

The operating point needs one methodological note. The open-source
timing analyzer for this part has no timing arcs for the MULT18X18 DSP
primitive, so static timing analysis cannot see the multiplier path. An
earlier build of the same design passed static timing at a reported
241~MHz and ran at 135~MHz, yet corrupted results under full-scale
operands: in an adversarial stress test with entries biased to the $\pm1$
extremes of Q1.15, only 1 of 100 rounds was bit-exact. The clock was
therefore lowered until the same stress test passes completely; the
108~MHz build is bit-exact in 500 of 500 rounds. The operating point is
stress-characterized rather than timing-analysis-derived, the logs of both
stress runs are in the repository, and the earlier version of this
manuscript reported the 135~MHz figures.

\begin{figure}[t]
  \centering
  \includegraphics[width=\columnwidth]{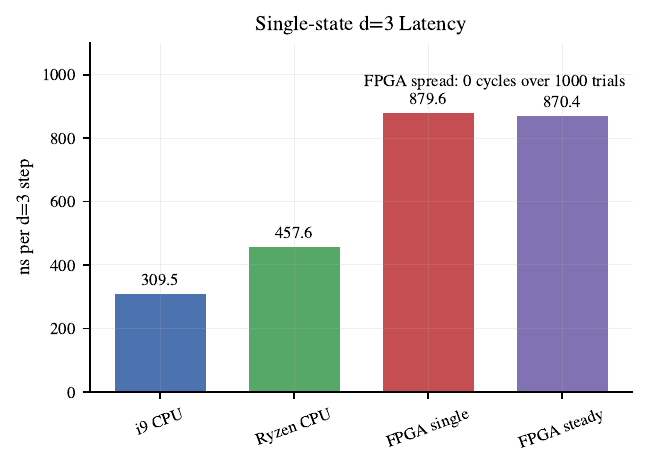}
  \caption{Single-state $d=3$ latency comparison. The FPGA is not the
  latency winner against the tested CPUs, but it is the
  deterministic-latency winner: the measured single-step count had zero
  spread over 1000 trials.}
  \label{fig:fpga_latency}
\end{figure}

\begin{table}[t]
  \centering
  \caption{Measured FPGA summary for the $d=3$ kernel.}
  \label{tab:fpga_summary}
  \small
  \resizebox{\columnwidth}{!}{\begin{tabular}{@{}lr@{}}
\toprule
quantity & value \\
\midrule
programmed clock & 108 MHz \\
single-step count & 95 cycles \\
single-step latency & 879.6 ns \\
long-run count & $94n + 1$ cycles \\
trial-to-trial spread & 0 cycles over 1000 trials \\
LUT4 usage & 1035 / 20736 \\
DSP usage & 4 / 48 \\
\bottomrule
\end{tabular}
}
\end{table}

\begin{figure*}[t]
  \centering
  \includegraphics[width=\textwidth]{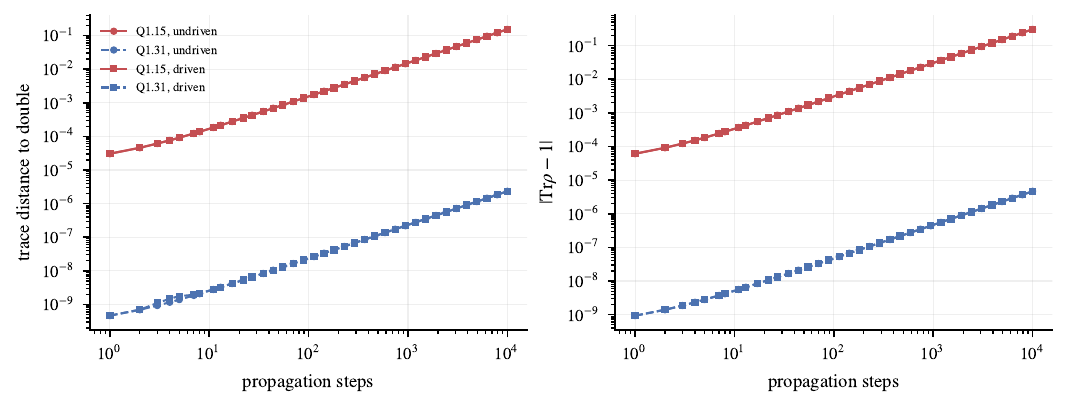}
  \caption{Fixed-point error growth for the $d=3$ propagator, undriven and
  driven ($0.05\,\sigma_x$), against the double-precision reference. Left:
  trace distance. Right: trace loss. Q1.15 loses $6.1\times10^{-5}$ of trace per step, two Q1.15
  least-significant bits (LSB), through floor truncation. Q1.31 stays below
  $10^{-5}$ over $10^4$ steps.}
  \label{fig:fixed_point_drift}
\end{figure*}

\begin{table}[t]
  \centering
  \caption{The same drift measured on the board. The Tang Nano ran $N$
  steps on chip from $|0\rangle\langle 0|$ and returned the Q1.15 state.
  "board = model" is bit-exact agreement with the software Q1.15 model of
  Figure~\ref{fig:fixed_point_drift}. Errors are against the double-precision
  reference. $p_1$ is the population of $|1\rangle$.}
  \label{tab:fpga_drift_hw}
  \scriptsize
  \resizebox{\columnwidth}{!}{\begin{tabular}{@{}lrlrrr@{}}
\toprule
case & steps & board = model & trace distance & $|\mathrm{Tr}\rho-1|$ & $|\Delta p_1|$ \\
\midrule
driven & 1 & yes & 3.05e-05 & 6.10e-05 & 2.79e-05 \\
driven & 10 & yes & 1.68e-04 & 3.36e-04 & 1.49e-04 \\
driven & 100 & yes & 1.54e-03 & 3.08e-03 & 8.77e-04 \\
driven & 1000 & yes & 1.53e-02 & 3.05e-02 & 1.33e-03 \\
driven & 10000 & yes & 1.53e-01 & 3.05e-01 & 2.00e-03 \\
undriven & 1 & yes & 3.05e-05 & 6.10e-05 & 0.00e+00 \\
undriven & 10 & yes & 1.68e-04 & 3.36e-04 & 0.00e+00 \\
undriven & 100 & yes & 1.54e-03 & 3.08e-03 & 0.00e+00 \\
undriven & 1000 & yes & 1.53e-02 & 3.05e-02 & 0.00e+00 \\
undriven & 10000 & yes & 1.53e-01 & 3.05e-01 & 0.00e+00 \\
\bottomrule
\end{tabular}
}
\end{table}

This FPGA result is not a throughput result. The best tested CPU
single-state latencies at $d=3$ were faster than the FPGA, and the best
batched CPU throughput was far faster. The value of the FPGA is
deterministic fixed latency inside hardware. The evidence supports exactly
that: one fixed-point kernel at one size, with cycle-exact repeatability
on chip. UART wall time is milliseconds and is dominated by transport, so
it does not enter the comparison. The accuracy side of the fixed-point
design is measured rather than assumed: Figure~\ref{fig:fixed_point_drift}
gives the drift of the Q1.15 arithmetic against double precision, and
Table~\ref{tab:fpga_drift_hw} shows the board reproducing the software
Q1.15 model bit for bit from 1 to $10^4$ on-chip steps, so the drift curve
is a property of the deployed hardware, not only of a simulation of it.

\subsection{Application Benchmark: Fast Propagation Does Not Guarantee Fast GRAPE}
\label{sec:grape}

Fast propagation does not guarantee fast GRAPE, and the $d=27$ case makes
that plain. With the aligned physical model of
Section~\ref{sec:protocol}, the C path is 101$\times$ faster than
released QuTiP \texttt{mesolve} on the i9 at $d=3$ and 224$\times$ on the
Ryzen, and still wins at $d=9$ by 2.8$\times$ and 3.4$\times$. At
$d=27$ the result reverses: \texttt{mesolve} wins by 2.3$\times$ on the
i9 and 4.9$\times$ on the Ryzen. The algorithm-matched QuTiP
\texttt{expm} baseline separates the two causes. Against it the C path
wins 39$\times$ and 89$\times$ at $d=3$ and loses only 1.2$\times$ and
2.0$\times$ at $d=27$, so most of the $d=27$ gap against
\texttt{mesolve} is algorithmic: an adaptive integrator crosses each
10-unit segment in fewer effective operations than twenty fixed propagator
applications plus one dense $729\times729$ matrix exponential per
segment. One asymmetry qualifies the $d=27$ comparison. The C path is
single-threaded, while QuTiP's BLAS uses every core by default. With BLAS
pinned to one thread (Table~\ref{tab:qutip_thread_sensitivity}) the QuTiP
\texttt{expm} path takes $11.6$~s at $d=27$ and the C path wins by
$1.7\times$, and the \texttt{mesolve} margin narrows to $2.1\times$. The
default-threading numbers stay primary because they are what an unmodified
installation delivers, but the deficit against the matched algorithm is a
threading effect, and the deficit against \texttt{mesolve} is partly one.

\begin{figure}[t]
  \centering
  \includegraphics[width=\columnwidth]{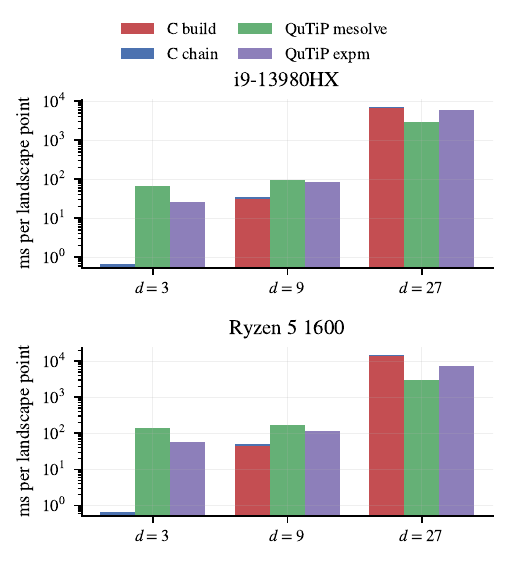}
  \caption{GRAPE-style landscape-point cost. The C path is separated into
  propagator construction and propagation chaining. The sign change at
  $d=27$ is not a contradiction of the propagation-kernel result; it shows
  that propagator construction becomes the dominant term once the generator
  is large enough.}
  \label{fig:grape_breakdown}
\end{figure}

\begin{table}[t]
  \centering
  \caption{End-to-end GRAPE-style benchmark: the C path against released
  QuTiP 5.2.3 \texttt{mesolve} (adaptive integration) and against QuTiP
  building and applying propagators via \texttt{expm} (algorithm-matched).
  Medians over trials, aligned physical model on all paths.}
  \label{tab:grape_summary}
  \scriptsize
  \resizebox{\columnwidth}{!}{\begin{tabular}{@{}llrrlrl@{}}
\toprule
host & $d$ & C (ms) & QuTiP mesolve (ms) & vs mesolve & QuTiP expm (ms) & vs expm \\
\midrule
i9 & $d=3$ & 0.677 & 68.607 & C ($101\times$) & 26.187 & C ($39\times$) \\
i9 & $d=9$ & 34.533 & 96.217 & C ($2.8\times$) & 83.379 & C ($2.4\times$) \\
i9 & $d=27$ & 6952.744 & 2995.333 & QuTiP ($2.3\times$) & 5963.345 & QuTiP ($1.2\times$) \\
Ryzen & $d=3$ & 0.639 & 143.428 & C ($224\times$) & 56.919 & C ($89\times$) \\
Ryzen & $d=9$ & 50.558 & 170.988 & C ($3.4\times$) & 118.981 & C ($2.4\times$) \\
Ryzen & $d=27$ & 14511.769 & 2937.072 & QuTiP ($4.9\times$) & 7261.275 & QuTiP ($2.0\times$) \\
\bottomrule
\end{tabular}
}
\end{table}

\begin{figure}[t]
  \centering
  \includegraphics[width=\columnwidth]{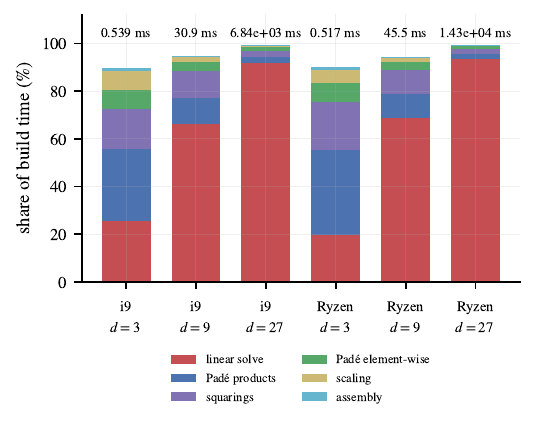}
  \caption{Where the propagator-build time goes. Each bar is one (host, $d$)
  pair, split by stage of the scaling-and-squaring Pad\'e build. The total
  build time per landscape point is printed above each bar.}
  \label{fig:grape_build_breakdown}
\end{figure}

\begin{table}[t]
  \centering
  \caption{Build-time breakdown per landscape point. $s$ is the number of
  squarings per segment propagator.}
  \label{tab:grape_build_breakdown}
  \scriptsize
  \resizebox{\columnwidth}{!}{\begin{tabular}{@{}llrrrrrr@{}}
\toprule
host & $d$ & build (ms) & solve (\%) & Pad\'e products (\%) & squarings (\%) & other (\%) & $s$ per segment \\
\midrule
i9 & $d=3$ & 0.5 & 25.7 & 30.0 & 17.0 & 27.2 & 3 \\
i9 & $d=9$ & 30.9 & 66.2 & 11.2 & 11.0 & 11.6 & 5 \\
i9 & $d=27$ & 6837.0 & 91.7 & 2.7 & 2.4 & 3.2 & 6 \\
Ryzen & $d=3$ & 0.5 & 19.8 & 35.5 & 20.1 & 24.6 & 3 \\
Ryzen & $d=9$ & 45.5 & 68.6 & 10.3 & 10.1 & 10.9 & 5 \\
Ryzen & $d=27$ & 14311.2 & 93.7 & 1.9 & 2.1 & 2.3 & 6 \\
\bottomrule
\end{tabular}
}
\end{table}

\begin{table}[t]
  \centering
  \caption{QuTiP timing sensitivity to BLAS threading on the i9. Default is
  the as-installed configuration. Pinned is single-threaded BLAS on the
  P-core set. Ranges are across trials.}
  \label{tab:qutip_thread_sensitivity}
  \scriptsize
  \resizebox{\columnwidth}{!}{\begin{tabular}{@{}llrrrr@{}}
\toprule
QuTiP path & $d$ & default median (ms) & default range & pinned median (ms) & pinned range \\
\midrule
mesolve & $d=3$ & 68.6 & 66.8--82.9 & 64.9 & 63.8--80.2 \\
mesolve & $d=9$ & 96.2 & 95.9--96.9 & 67.0 & 66.8--67.4 \\
mesolve & $d=27$ & 2995.3 & 1798.8--3431.1 & 3248.4 & 3242.7--3265.6 \\
expm & $d=3$ & 26.2 & 25.4--43.9 & 17.3 & 16.7--17.7 \\
expm & $d=9$ & 83.4 & 80.6--96.2 & 68.7 & 68.5--68.8 \\
expm & $d=27$ & 5963.3 & 4155.7--8777.8 & 11582.6 & 11413.1--11616.1 \\
\bottomrule
\end{tabular}
}
\end{table}

Figure~\ref{fig:grape_breakdown} shows where the time goes. The
propagation chain is not the problem at $d=27$; dense propagator
construction dominates the landscape-point cost. Figure~
\ref{fig:grape_build_breakdown} and Table~\ref{tab:grape_build_breakdown}
resolve the build term itself: the LU solve of the Pad\'e denominator,
line~6 of Algorithm~\ref{alg:build}, takes $92\%$ of the build on the i9
and $94\%$ on the Ryzen at $d=27$, while the Pad\'e products and the six
squarings stay small because the zero-skipping product exploits the
sparsity of the powers of $\mathcal{L}\Delta t$ and the solve cannot. The
measured picture is therefore consistent across levels: a real kernel
advantage, a real application advantage at $d=3$ and $d=9$, and an
application deficit at $d=27$ that is located in one dense
$\mathcal{O}(n^3)$ solve.

\FloatBarrier
\section{Discussion}

Three operating regimes emerge from the measured data.

First, CPU execution is the correct default for tiny dense work and for any
control loop where the available parallelism is limited. The CPU does not
win here because it has the highest theoretical peak. It wins because the
problem is small enough that low overhead and cache residency matter more
than raw peak arithmetic.

Second, the GPU story depends on which timing question is being asked. If
the question is whether a resident device kernel can outrun the CPU once
there is enough batched work, the answer is yes for the larger measured
cases. If the question is whether the full host-visible platform path beats
the CPU in the measured range, the answer is no, and the resident-$P$
measurements locate the difference in the propagator re-upload and
allocation of the host-visible path. The three models answer three
deployment questions and are reported separately.

Third, the FPGA occupies a different corner of the design space. It is not a
throughput replacement for CPU simulation. It is a deterministic-latency
result for an embedded control path. That distinction matters for
superconducting-qubit feedback electronics, where fixed response and jitter
can matter as much as bulk throughput~\cite{Gebauer2020,Yang2021,Caune2024}.

The GRAPE benchmark then gives the practical recommendation. For the
current code base, the next useful optimization target at $d=27$ is not
another round of propagation-loop tuning and not propagator construction in
general: the breakdown of Table~\ref{tab:grape_build_breakdown} narrows it
to the dense LU solve of the Pad\'e denominator, which takes over $90\%$
of the build. Three follow-on directions attack that line. A blocked or
multithreaded LU brings standard dense-linear-algebra techniques to the one
dense $\mathcal{O}(n^3)$ step this code still runs naively. Deeper
control-affine reuse around $\mathcal{L} = \mathcal{L}_0 + \Omega
\mathcal{L}_{\mathrm{drv}} + \mathcal{L}_{\mathrm{diss}}$ would let
closely related controls avoid full rebuilds. Replacing the dense
propagator with the action of $e^{\mathcal{L}\Delta t}$ on the state via
a Krylov method~\cite{Sidje1998,AlMohy2011} removes the solve entirely,
at the cost of leaving the fixed-propagator regime that the rest of this
paper measures. The present data do not choose among the three, but they
isolate the line any of them must beat.

\FloatBarrier
\section{Conclusion}

This study measured small dense Lindblad propagation at the sizes that
transmon gate simulation uses, a regime where cache placement, launch
overhead, and deterministic latency change the answer. For the measured
grid, three sizes, two CPUs, two GPUs, one $d=3$ fixed-point FPGA design,
and one GRAPE-style benchmark, the results split into operating regimes,
and each measured cost is within a stated factor of an idealized cost built
from separately measured ceilings.

CPU execution is the default for very small low-overhead work. GPU
execution beats the CPU in kernel-only timing before it does in
host-visible timing, and the resident-$P$ model shows that most of the
difference is propagator re-upload and allocation, so all three timing
models carry information. The Tang Nano FPGA establishes a
deterministic-latency $d=3$ operating point, 95 cycles at a
stress-characterized 108~MHz with zero spread over 1000 trials and
board-verified Q1.15 drift, but it is not a throughput winner. At the
application level, the C GRAPE benchmark wins by two orders of magnitude at
$d=3$, still wins at $d=9$, and loses at $d=27$, where the dense LU solve
inside the Pad\'e build takes over $90\%$ of the cost.

The practical conclusion for this grid is a division of labor: CPU for tiny
dense work and low overhead, GPU for sufficiently batched propagation once
host costs are amortized, and FPGA where deterministic embedded latency is
the requirement. For the largest measured size, the next method step is the
one line the breakdown isolates: the dense solve inside propagator
construction, not propagation itself.

\section*{Data and Code Availability}

All benchmark code, raw timing data, and figure-generation scripts used
in this study are available in the \texttt{lindblad-bench} repository at
\url{https://github.com/rylanmalarchick/lindblad-bench}. It also contains the
CSV and log artifacts used to generate the reported tables and figures.

\section*{Declaration of Generative AI and AI-Assisted Technologies in the Writing Process}

OpenAI Codex and Anthropic Claude Code were used to assist with software
development, benchmark automation, figure-generation scripts, and language
editing. The author reviewed and edited all generated material and takes full
responsibility for the content of the publication.

\bibliographystyle{elsarticle-num}
\bibliography{refs}

\end{document}